\documentclass[aps,prl,superscriptaddress,twocolumn,floatfix]{revtex4-1}
\usepackage{graphicx}
\usepackage{amssymb}
\usepackage{amsfonts}
\usepackage{amsmath}

\begin{document}

\title{Type I superconductivity in the Dirac semimetal PdTe$_2$}

\author{H. Leng} \email{h.leng@uva.nl}\affiliation{Van der Waals - Zeeman Institute, University of Amsterdam, Science Park 904, 1098 XH Amsterdam, The Netherlands}
\author{C. Paulsen} \affiliation{Institut N\'{e}el, CNRS \& Universit\'{e} Grenoble Alpes, BP 166, 38042 Grenoble Cedex 9, France}
\author{Y. K. Huang} \affiliation{Van der Waals - Zeeman Institute, University of Amsterdam, Science Park 904, 1098 XH Amsterdam, The Netherlands}
\author{A. de Visser} \email{a.devisser@uva.nl} \affiliation{Van der Waals - Zeeman Institute, University of Amsterdam, Science Park 904, 1098 XH Amsterdam, The Netherlands}

\date{\today}

\begin{abstract}
The superconductor PdTe$_2$ was recently classified as a Type II Dirac semimetal, and advocated to be an improved platform for topological superconductivity. Here we report magnetic and transport measurements conducted to determine the nature of the superconducting phase. Surprisingly, we find that PdTe$_2$ is a Type I superconductor with $T_c = 1.64$~K and a critical field $\mu_0 H_c (0) = 13.6$~mT. Our crystals also exhibit the intermediate state as demonstrated by the differential paramagnetic effect. For $H > H_c$ we observe superconductivity of the surface sheath. This calls for a close examination of superconductivity in PdTe$_2$ in view of the presence of topological surface states.
\end{abstract}

\maketitle

Recently the transition metal dichalcogenide PdTe$_2$ was reported to be a Type II Dirac semimetal~\cite{Noh2017,Fei2017,Soluyanov2017}. Topological Dirac semimetals form a new class of topological materials, where non-trivial surface states arise due to the topology of the bulk band structure (for recent reviews see~\cite{Chiu2016,Yan&Felser2017,Armitage2017}). Dirac semimetals are the 3D analog of graphene and have a cone-shaped linear energy dispersion around the Dirac point with massless fermions~\cite{Young2012}. The bands have a double degeneracy that can be lifted by a magnetic field resulting in a pair of Dirac cones. In the closely related class of Weyl semimetals the degeneracy is naturally lifted by breaking time reversal and/or inversion symmetry~\cite{Wan2011}. The set of Dirac cones can give rise to distinct properties, such as Fermi arcs at the surface, quantum anomalous Hall effect and chiral magnetotransport~\cite{Yan&Felser2017,Armitage2017}. Type I Dirac semimetals are like graphene and the valence and conduction bands meet at the Dirac point and Lorentz invariance is obeyed. In Type II Dirac semimetals an extra momentum dependent term in the Hamiltonian breaks Lorentz invariance~\cite{Soluyanov2015,Huang2016,Yan2017}. This can be accomplished by tilting the Dirac cone, where the Dirac point is now the touching point of the electron and hole pockets. This gives rise to a number of new physical phenomena, such as an angle dependent chiral anomaly and topological Lifshitz transitions~\cite{Yan&Felser2017,Armitage2017}.

Superconductivity in PdTe$_2$ with a transition temperature $T_c$ of 1.5~K was discovered in 1961~\cite{Guggenheim1961}. The recent detection of topological features in the band structure raises the question whether superconductivity has also a topological nature~\cite{Liu2015a,Fei2017,Noh2017}. Notably, it has been advocated that PdTe$_2$ is an improved platform for topological superconductivity~\cite{Fei2017}. Topological superconductors attract much attention because they are predicted to host protected Majorana zero modes at their surface (for recent reviews see~\cite{Sato&Fujimori2016,Sato&Ando2017}). This offers a unique design route to produce future devices for topological quantum computation. Unfortunately, the number of materials in which topological superconductivity has been realized - or is under debate - is very small~\cite{Sato&Ando2017}. Majorana modes, that appear as gapless nodes in the bulk superconducting gap, are in general not stable in a Type I Dirac semimetal~\cite{Sato&Ando2017}. However, in a Type II semimetal the  situation is different because of the tilted dispersion. Moreover, the abundancy of states in the electron and hole pockets near the Type II Dirac point favours a larger carrier concentration and superconductivity~\cite{Fei2017}.

Hitherto, the superconducting state of PdTe$_2$ has not been studied in detail. The early determination of $T_c$ by Guggenheim \textit{et al.}~\cite{Guggenheim1961} was confirmed by others with $T_c$ values ranging from 1.7 to 2.0~K~\cite{Kjekshus&Pearson1965,Raub1965,Wang2016,Fei2017}. Fei \textit{et al}.~\cite{Fei2017} investigated the depression of $T_c$ in magnetic field and reported an anomalous upward curvature of the upper critical field $H_{c2}(T)$ with $\mu_0 H_{c2} = 0.32$~T for $T\rightarrow 0$. In view of the proposed topological nature of the superconducting state~\cite{Liu2015a,Fei2017,Noh2017} an in depth characterization of the superconducting phase is a matter of great urgency. Here we report magnetic and transport measurements on single crystals that unambiguously show PdTe$_2$ is a Type I superconductor. This makes PdTe$_2$ the first topological material where superconductivity is of Type I. This is a surprising results, also because the number of known binary and ternary systems with Type I superconductivity is very small (see for instance Refs.~\cite{Zhao2012,Sun2016,Kimura2016} and references therein). Our crystals also show enhanced superconductivity of the surface sheath in fields exceeding the critical field $H_c$. The surface superconductivity does not obey the standard Saint-James - de Gennes behavior with critical field $H_{c3} = 1.69 \times H_c$~\cite{Saint-James&deGennes1963}. We discuss these unusual results in view of the presence of topological surface states~\cite{Liu2015a,Noh2017}.

PdTe$_2$ crystallizes in the trigonal CdI$_2$ structure (space group P$\bar{3}$m1)~\cite{Thomassen1929}. It belongs to the family of transition metal dichalcogenides, which is intensively studied because of the remarkable physical properties~{\cite{Manzeli2017}. Its normal-state electronic properties have been investigated in the 1970s by quantum oscillation experiments and band structure calculations~\cite{Myron1974,Dunsworth1975,Jan&Skriver1977}. The topological nature of the electronic band structure was reported recently~\cite{Liu2015a,Fei2017,Noh2017}. Notable angle resolved photoemission spectroscopy (ARPES) combined with \textit{ab initio} band structure revealed PdTe$_2$ is a Type II Dirac semimetal~\cite{Noh2017}, which finds further support in a non-trivial Berry phase originating from a hole pocket formed by a tilted Dirac cone~\cite{Fei2017}. The fundamental electronic properties of PdTe$_2$ were revisited recently by transport, magnetic and thermal measurements~\cite{Hooda&Yadav2017}.

For our study of the superconducting properties of PdTe$_2$ we prepared a single crystal by a modified Bridgeman technique~\cite{Lyons1976}. Powder X-ray diffraction confirmed the CdI$_2$ structure. Scanning Electron Microscopy (SEM) with Energy Dispersive X-ray (EDX) spectroscopy showed the proper 1:2 stoichiometry within the experimental resolution of 0.5~\% (see Supplemental Material (SM)). Laue back-scattering was used to orient the crystal. Single crystalline bars, typically a few mm long, were cut along the crystallographic $a$-axis by means of a scalpel blade and/or spark erosion. Standard  four-point resistance measurements were carried out in a Physical Property Measurement System (Quantum Design) at temperatures down to 2~K and in a helium-3 refrigerator (Heliox, Oxford Instruments) down to 0.3~K. Dc-magnetization, $M(T,H)$, and ac-susceptibility, $\chi _{ac}(T,H)$, measurements were made using a low field SQUID magnetometer developed at the N\'{e}el Institute. The magnetometer is equipped with a miniature dilution refrigerator making possible absolute value measurements by the extraction technique. A MuMetal and superconducting shield combination results in a residual field of a few milliOersted at the sample position when cooled. As regards $\chi_{ac}$, the in-phase, $\chi_{ac} ^{\prime}$, and out-of-phase, $\chi_{ac} ^{\prime \prime}$, signals were measured in driving fields $\mu_0 H_{ac} = 0.0005 - 0.25$~mT with low frequencies $f_{ac}=2.3-13$~Hz.

In Fig.~\ref{figure:M_H_s2} we show the dc-magnetization as a function of the applied field $H_a$ in the temperature range $0.31-1.50$~K. The $M(H_a)$-curves follow the behavior of a Type I superconductor with a Meissner phase up to $\mu_0 H_a = 12$~mT and the intermediate state for $12 < \mu_0H_c < 13.6$~mT, where $H_c$ is the critical field. The large value of the measured initial slope $\chi_m = dM/dH_a =\chi (1+N\chi) = -1.13$ is in agreement with bulk superconductivity. Here $N$ is the demagnetization factor and $\chi = -1$ the ideal susceptibility~\cite{Poole2007}. From the initial slope we calculate $N=0.12$ which is close to the estimated value $\sim 0.10$ based on the sample shape (see SM). We remark the rounding of the curves is due to the non-uniform magnetization at the sample edges. However, a clear kink and tail is observed in the data just above $H_c$ (see inset). We will return to this point later. We have determined $H_c(T)$ by extrapolating the idealized linear $M(H_a)$-curves to $M=0$, as shown by the dash-dotted line for $T=0.31$~K in Fig.~\ref{figure:M_H_s2}. The critical field follows the standard BCS quadratic temparature variation $H_c(T) = H_c(0)[1-(T/T_c)^2]$, with $\mu_0 H_c = 13.6$~mT and $T_c$ = 1.64~K, see Fig.~\ref{figure:PD_s2}.

\begin{figure}[ht]
\centering
\includegraphics[width=8cm]{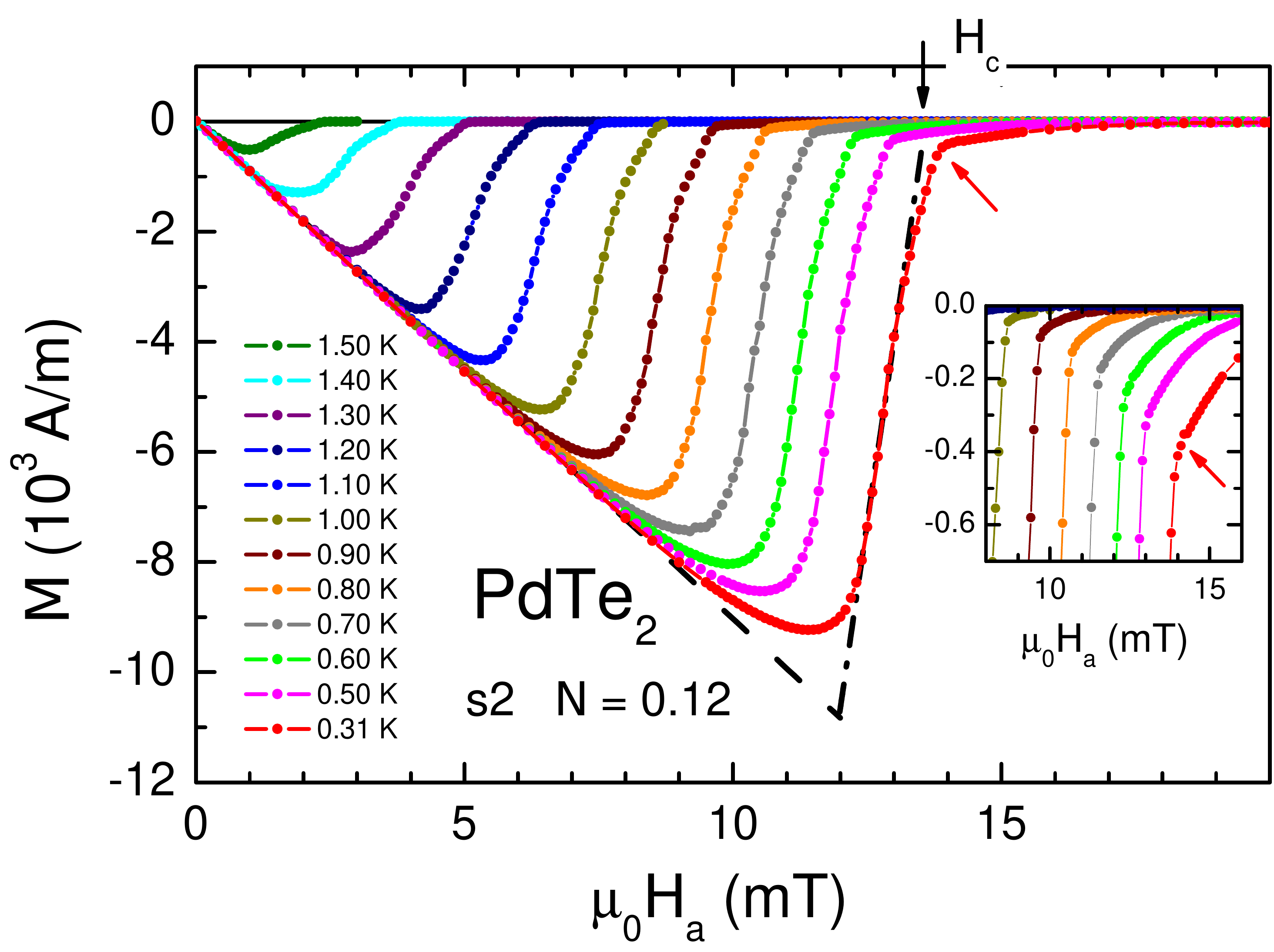}
\caption{Dc-magnetization per unit volume (S.I. units) as a function of applied field for PdTe$_2$ at temperatures from 0.31~K (right) to 1.50~K (left) as indicated. The initial slope $\chi_m = dM/dH_a$ accounts for a superconducting sample volume of 100~\% with $N=0.12$ (dashed line). The dash-dotted line indicates the idealized $M(H_a)$-curve with slope $1/N$ in the intermediate state at $T=0.31$~K. The black arrow indicates $H_c$ at $T=0.31$~K. The red arrow points to a kink and start of a tail in $M(H_a)$. Inset: Zoom of the kink-feature at a few selected temperatures. }
\label{figure:M_H_s2}
\end{figure}

\begin{figure}[ht]
\centering
\includegraphics[width=8cm]{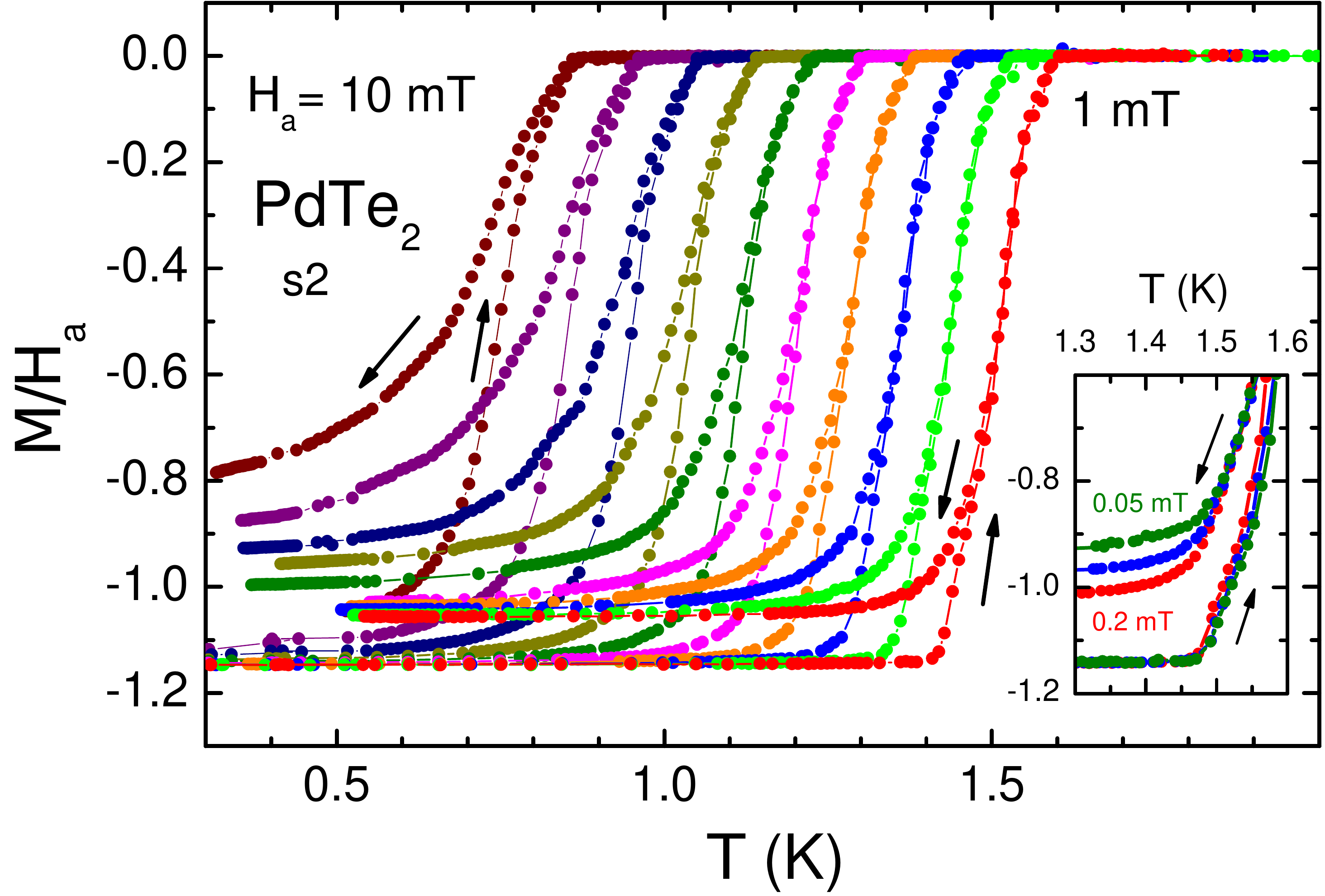}
\caption{Dc-susceptibility, $M/H_a$, in S.I. units, as a function of temperature in fields $\mu_0 H_a$ from 1~mT (right) to 10~mT (left) in steps of 1~mT. Data are taken after cooling in zero field (ZFC) and field cooled (FC) as shown by the arrows. Inset: Part of the ZFC-FC curves in applied fields of 0.2 (red), 0.1 (blue) and 0.05~mT (green). }
\label{figure:X_T_s2}
\end{figure}

The temperature variation of the dc-susceptibility, $\chi_{dc}(T)$, in applied fields $\leq 10$~mT is shown in Fig.~\ref{figure:X_T_s2}. The data are taken after cooling in zero field (ZFC) and field cooled (FC). The FC data at low applied dc-fields ($\mu_0 H_a= 1$~mT) demonstrate a large Meissner effect with a flux expulsion of 93~\%. Ac-susceptibility measurements in an ac-driving field $\mu_0 H_{ac} = 0.25$~mT for dc-fields up to 10~mT are reported in Fig.~\ref{figure:chi_ac_s2}a,b. At low temperatures $\chi_{ac} ^{\prime}$ shows a full superconducting screening signal. Upon increasing the temperature  $\chi_{ac} ^{\prime}$ does not show the usual smooth increase to zero. Instead the signal becomes positive and shows a large peak before the normal state is reached at $T_c$. This is known as the differential paramagnetic effect (DPE)~\cite{Hein&Falge1961}. It results from the positive $\partial M/ \partial H_a$ below $H_c$ in the intermediate state (see Fig.~\ref{figure:M_H_s2}), $i.e.$ in between $(1-N)H_c$ and $H_c$, and has been observed in other Type I superconductors as well~\cite{Zhao2012,Kimura2016}. $H_c (T)$-data points extracted from the dc and ac-susceptibility data in fixed fields have been collected in Fig.~\ref{figure:PD_s2} as well.

The dc-magnetization, the ac-susceptibility with DPE and the extracted $T^2$-variation of $H_c$, all provide solid evidence PdTe$_2$ is a Type I superconductor. This tells us the Ginzburg-Landau parameter $\kappa = \lambda / \xi < 1/\sqrt{2}$. An estimate for the magnetic penetration depth, $\lambda$, can be obtained using the London equation $\lambda = (m^*/\mu_0 n_s e^2)^{1/2}$, where $m^*$ is the effective mass, $n_s$ the superfluid density and $e$ the elementary charge. With a carrier density $n=3.8 \times 10^{28}$~m$^{-3}$~\cite{Hooda&Yadav2017}, and $ m^* \approx 0.3m_e$~\cite{Dunsworth1975,Wang2016} (here we use an average value $ m^*$ and $m_e$ is the free electron mass) we arrive at $\lambda \sim 15$~nm. A value for the superconducting coherence length, $\xi$, can be derived from the Ginzburg-Landau relation $\xi=\Phi_0/(2\sqrt 2 \pi \mu_0 H_c \lambda)$~\cite{Tinkham1996}, here $\Phi_0$ is the flux quantum. With the measured value $H_c (0) = 13.6$~mT we obtain $\xi \approx 114$~nm, and calculate $\kappa \approx 0.13$. We remark that realistic errors margins in the values of $n$ and $m^*$ will not affect the result $\kappa < 1/\sqrt 2$. Since $-\mu_0 H_c^2 /2$ is the condensation energy per unit volume we can use thermodynamic relations to calculate $H_c$ from the step-size of the specific heat at $T_c$ using the relation $\Delta C |_{T_c} = 4 \mu_0 H_c(0)^2 /T_c =1.43 \times \gamma T_c$~\cite{Poole2007}, assuming PdTe$_2$ is a weak coupling BCS superconductor~\cite{Kudo2016}. Here $\gamma$ is the Sommerfeld coefficient. With the experimental value $\gamma = 138$~J/K$^2$m$^3$~\cite{Hooda&Yadav2017,Kudo2016} (the molar volume is $4.34 \times 10^{-5}$~m$^3$/mol), we calculate $\mu_0 H_c (0)$ = 12.6~mT, which is close to the measured value reported in Fig.~\ref{figure:PD_s2}.

\begin{figure}[ht]
\centering
\includegraphics[width=8.5cm]{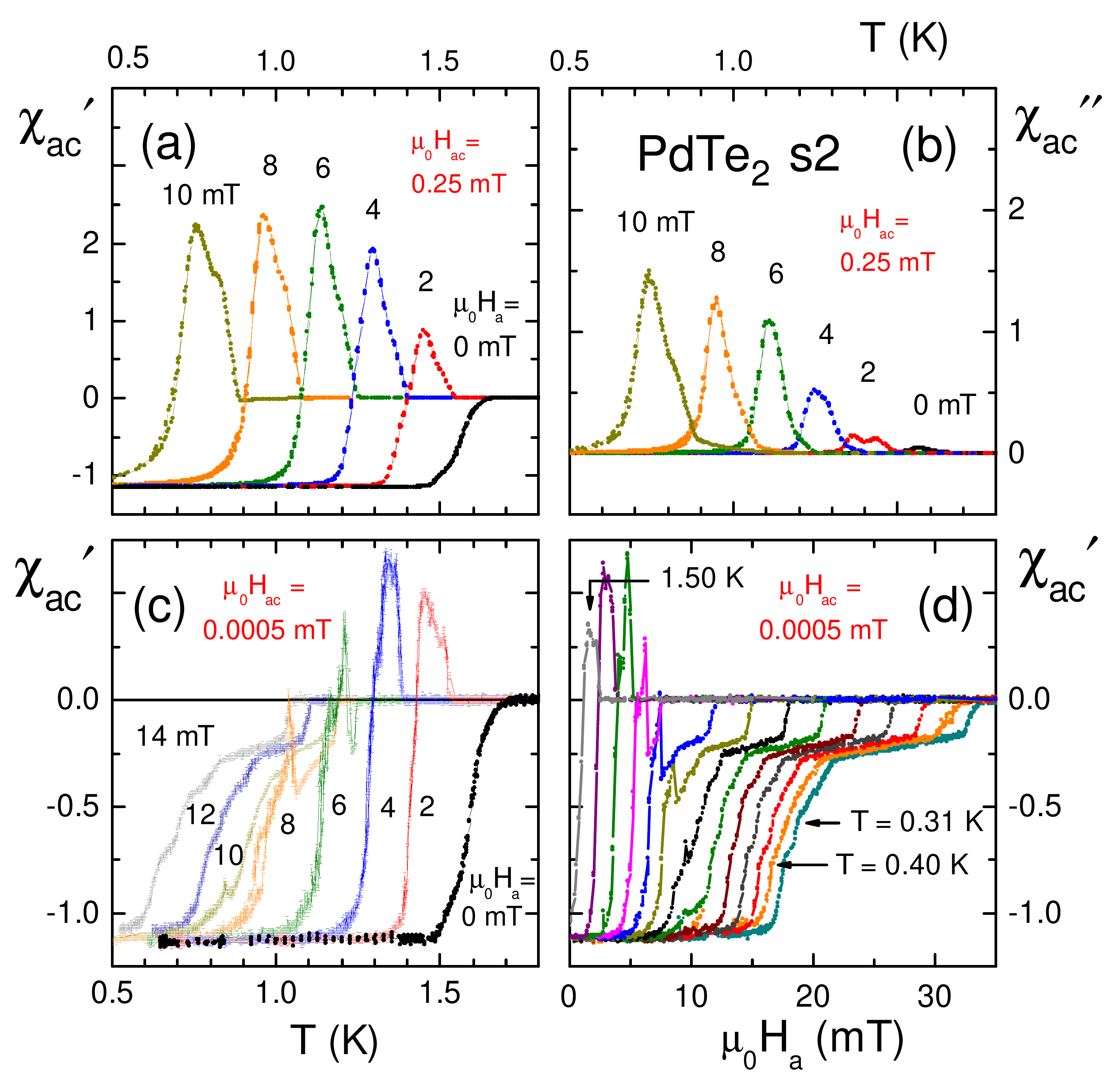}
\caption{Ac-susceptibility of PdTe$_2$. Upper panels (a) and (b): In phase and out-of-phase component of the ac-susceptibility for an ac-driving field $\mu_0 H_{ac}= 0.25$~mT. Data are taken in dc fields $\mu_0 H_a = 0-10$~mT, as indicated. The large peaks in $\chi _{ac} ^{\prime}$ when a dc field is applied are due to the differential paramagnetic effect. Lower panels: Ac-susceptibility in a small ac-driving field $\mu_0 H_{ac}=0.0005~$mT. Panel (c): As a function of temperature at dc-fields from 0 to 14~mT as indicated. Panel (d): As a function of applied field at a temperature of 0.31~K, and from 0.40 K to 1.50 K in steps of 0.1 K.}
\label{figure:chi_ac_s2}
\end{figure}

Having established that PdTe$_2$ is a bulk Type I superconductor, we next turn to superconductivity of the surface sheath. A close inspection of the $M(H)$ isotherms reported in Fig.~\ref{figure:M_H_s2} reveals a clear kink in the data close to $H_c$ and a long tail for $M(H) \rightarrow 0$ (see inset). Thus superconductivity survives above $H_c$. This is also most clearly observed in the ac-susceptibility data measured in a small driving field $\mu_0 H_{ac} = 0.0005$~mT reported in Fig.~\ref{figure:chi_ac_s2}c,d. For small fields ($\mu_0 H_a \leq 4$~mT) the $\chi _{ac} ^{\prime}(T)$-data (panel c) show the same behavior as reported in Fig.~\ref{figure:chi_ac_s2}a ($\mu_0 H_{ac} = 0.25$~mT). However, for $\mu_0 H_a \geq 6$~mT the DPE peak progressively reduces and screening persists even for fields exceeding $H_c$. The $\chi _{ac} ^{\prime}(H_a)$-data (panel d) show that at the lowest temperature (0.31~K) screening of the full superconducting volume takes place till $\sim$~17~mT. By further increasing $H_a$ the screened volume is reduced in a step-wise fashion, until finally at 33~mT the diamagnetic signal disappears completely. Since the $\chi _{ac} ^{\prime}(T,H_a)$-data show a full screening signal above $H_c$ this signal must come from the superconducting surface layer. This also explains why the large peak due to the DPE located just below $H_c$ becomes smaller and smaller with increasing applied field (panel c) or decreasing temperature (panel d): the bulk is screened by the surface layer~\cite{Weber&McEvoy1973}. The screening efficacy of the surface layer strongly depends on the amplitude of $H_{ac}$ (see SM). In Fig.~\ref{figure:chi_ac_s2}a $\mu_0 H_{ac} = 0.25$~mT and the screening is weak, while in Fig.~\ref{figure:chi_ac_s2}b $\mu_0 H_{ac} = 0.0005$~mT and the screening is large. It tells us flux pinning in the surface sheath is extremely weak and can be overcome by a driving field of typically 0.25~mT. The weak pinning at the surface of the crystal also explains why the FC dc-susceptibility  measured in very small dc-fields $\leq$~0.2~mT shows less flux expulsion than for fields $\geq 1.0$~mT (see inset Fig.~\ref{figure:X_T_s2}).

Next we present the superconducting phase diagram derived from the magnetic and transport measurements (Fig.~\ref{figure:PD_s2}). Superconductivity of the bulk is found below the $H_c$-phase line. The critical field of the surface layer $H_{c}^{s} (T)$ is identified from the data in Fig.~\ref{figure:chi_ac_s2}c,d by the field ($> H_c)$ at which $\chi _{ac} ^{\prime}(H)$ or $\chi _{ac} ^{\prime}(T)$ reaches zero. We remark that for the small amplitude ac-field, $\mu_0 H_{ac} = 0.0005$~mT, $H_{c}^{s} (T)$ is well defined due to the step-like feature when $\chi _{ac} ^{\prime} \rightarrow 0$. For larger amplitudes of $H_{ac}$ the step broadens (see SM). Obviously, $H_{c}^{s} (T)$ does not follow the standard relation for surface superconductivity $H_{c3} = 1.69 \times H_c$~\cite{Saint-James&deGennes1963}. Moreover, the extrapolation of $H_{c}^{s} (T)$ to $H \rightarrow 0$ reveals $T_c^s$ of the surface layer is 1.33~K, which is lower than the bulk $T_c$ (see Fig.~\ref{figure:PD_s2}). Here we fitted $H_{c}^{s} (T)$ to a quadratic temperature function, from which we infer $\mu_0 H_{c}^{s}(0)= 34.9$~mT. Remarkably, electrical resistance measurements for $H_a \parallel a$ on the same PdTe$_2$ crystal reveal superconductivity survives up to fields that are almost a factor 10 higher (see the right panel in Fig.~\ref{figure:PD_s2} and SM for details). The critical field determined by transport, $H_{c}^{R} (T)$, tracks the $H_c (T)$ curve for low fields (see SM), but increases rapidly below $\sim 1.3$~K. This temperature coincides, within the error bar, with $T_c^s$, which strongly suggests the transport experiment probes superconductivity of the surface layer as well. The $H_{c}^{R} (T)$-curve compares quite well with the standard Werthamer-Helfand-Hohenberg (WHH) expression for a weak-coupling spin-singlet superconductor in the clean limit~\cite{Werthamer1966} (see SM).

\begin{figure}[ht]
\centering
\includegraphics[width=8cm]{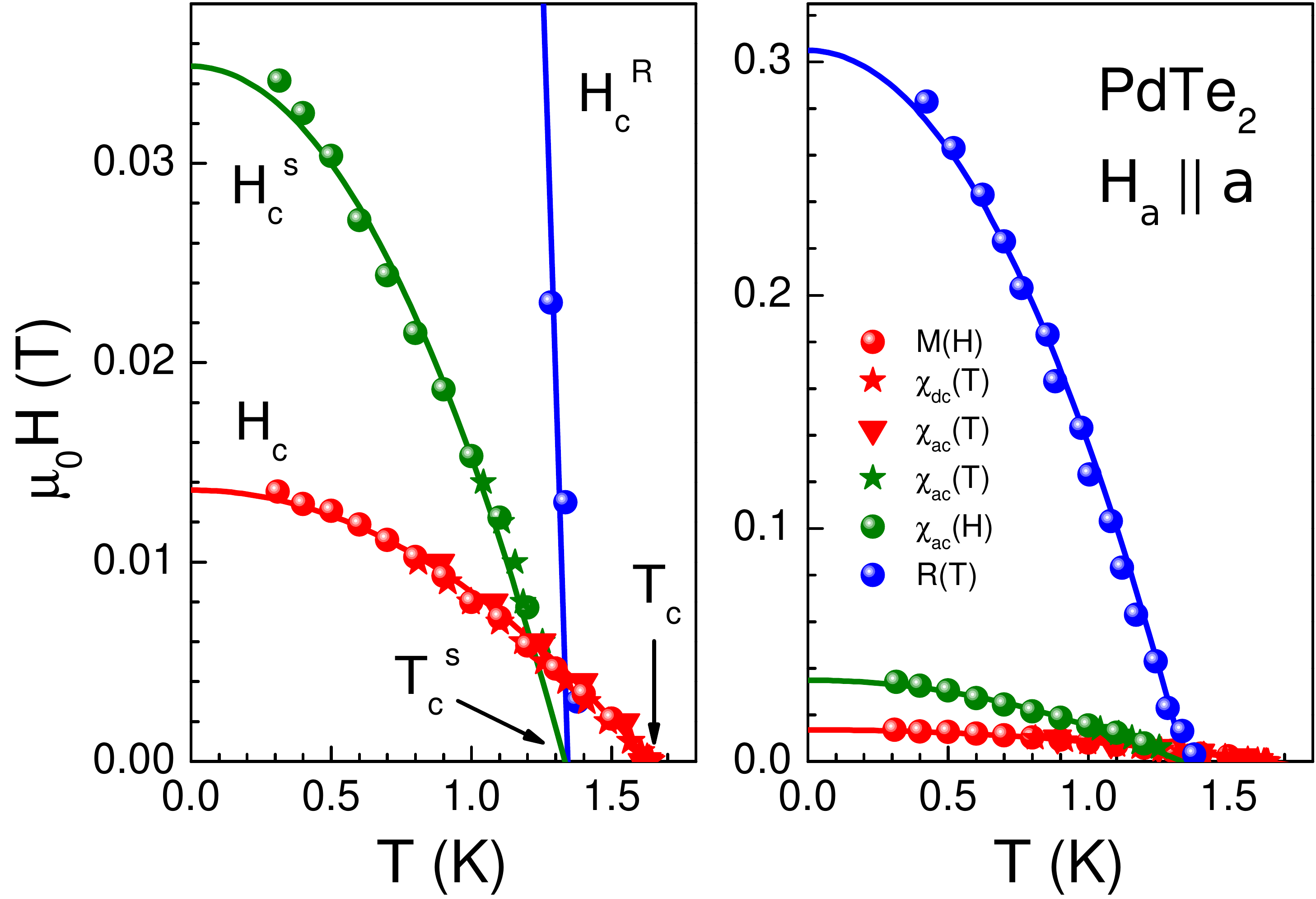}
\caption{Superconducting phase diagram of PdTe$_2$ for $H_a \parallel a$-axis. Bulk superconductivity is found below $H_c (T)$ as determined by dc-magnetization and $\chi _{ac} ^{\prime} $. The red line represents a fit to $H_c(T) = H_c(0)[1-(T/T_c)^2]$, with $\mu_0 H_c (0)= 13.6$~mT and $T_c$ = 1.64~K. Surface superconductivity is found below $H_c^s (T)$ as determined by $\chi _{ac} ^{\prime} $ for a small amplitude of $H_{ac}$ (see text). The green line represents a fit to $H_c^s(T) = H_c^s(0)[1-(T/T_c^s)^2]$, with $\mu_0 H_c^s (0)= 34.9$~mT and $T_c^s = $1.33~K. The blue symbols denote $H_{c}^{R} (T)$ and are taken from the superconducting transition measured by resistance. The blue line compares $H_{c}^{R} (T)$ with the WHH model curve (see text). }
\label{figure:PD_s2}
\end{figure}

The phase diagram with Type I superconductivity below $T_c= 1.64$~K and surface superconductivity below $T_c^s = 1.33$~K is at odds with the standard BCS behavior, but we stress it is a robust property of our PdTe$_2$ crystals. We have performed a number of checks. First of all SEM and EDX showed our crystals to have a homogeneous 1:2 composition and no foreign phases were detected (see SM). Secondly, and most importantly, after taking the $M$ and $\chi _{ac} ^{\prime}$ data we carefully polished the surfaces of the crystal and remeasured the magnetic properties with essentially the same results for the bulk and surface (see SM). This provides compelling evidence surface superconductivity is not due to an impurity phase on the surface. We emphasize the large critical field $H_{c}^{R}(T)$ measured by resistance is a robust property of our crystals as well. Resistance measurements for $B \parallel a^*$- and $c$-axis on the same crystal, as well as on other crystals, all show similarly enhanced values of $H_{c}^{R}(T)$ (see SM). The close to isotropic behavior for $B \parallel a$-, $a^*$- and $c$-axis indicates the superconducting transition in resistance is not due to filamentary superconductivity (see SM). Finally, we remark that Fei \textit{et al}.~\cite{Fei2017} reported a large critical field $\sim 0.32$~T for $T \rightarrow 0$ deduced from resistance data too.

The unusual superconducting phase diagram of PdTe$_2$ shows some similarities with the diagrams reported for the Type I superconductors LaRhSi$_3$~\cite{Kimura2016} and AuBe~\cite{Rebar2015}. For these materials also a surface critical field much larger than $H_c$ is found. However, in both case it was attributed to a field induced change from Type I to Type II/1 superconductivity below a conversion temperature $T^* < T_c$, which is possible when $\kappa$ is close to $1/\sqrt 2$~\cite{Auer&Ullmaier1973}. We remark that for PdTe$_2$ $\kappa = 0.13 < 1/\sqrt2$ and we did not find any evidence for a conversion from Type I to Type II/1. On the other hand, both LaRhSi$_3$ and AuBe have a noncentrosymmetric crystal structure. Theory predicts the lack of inversion symmetry can possibly give rise to exotic superconducting properties due to the mixing of spin-singlet and triplet order parameters~\cite{Yip2013}, as well as to unusual surface states. This possibly explains the measured critical fields are much larger than $H_c$.

The structure of superconducting states in Dirac semimetals was recently investigated by theoretical work~\cite{Hashimoto2016,Kobayashi2015,Fei2017,Bednik2015}. Depending on the different pairing potentials, topological odd-parity superconductivity in the bulk with gap nodes is a possibility. Since we find that PdTe$_2$ is a conventional BCS superconductor, such a scenario is most likely ruled out. On the other hand, ARPES measurements in the normal state reveal the presence of a topological surface state~\cite{Liu2015a,Noh2017}. Possibly, a superconducting gap opens in this topological surface state at $T_c ^s$, below $T_c$ of the bulk. Since, superconductivity of the surface layer, with two critical fields $H_c^s$ and $H_c^R$, does not follow the standard BCS behavior, we speculate it could have a topological nature. This calls for an in depth examination of superconductivity in PdTe$_2$, by \textit{e.g. } scanning tunneling probe techniques.

In summary, we have investigated the superconducting properties of the compound PdTe$_2$ that was recently reported to be a Type II Dirac semimetal. Dc-magnetization and ac-susceptibility measurements clearly show PdTe$_2$ is a Type I superconductor with $T_c = 1.64$~K and a critical field $\mu_0 H_c (0) = 13.6$~mT. Our crystals also show the intermediate state as is demonstrated by the differential paramagnetic effect observed in the ac-susceptibility. In addition, superconductivity of the surface layer is found below $T_c^s = 1.33$~K $< T_c$. It persists up to $\mu_0 H_c^s (0) = 34.9$~mT and does not follow the standard Saint-James - de Gennes behavior. Resistance data point to an even larger critical field for the surface layer $H_c^R(0) \approx 0.30$~T. PdTe$_2$ is the first topological material with Type I superconductivity. Together with the unusual superconducting phase diagram, this calls for a close examination of superconductivity in PdTe$_2$, especially in view of the existence of topological surface states.

\begin{acknowledgements}

We thank C.M. van Kats and A. Lof for assistance with the SEM/EDX crystal characterization and J. Debray for help with polishing crystal S2. AdV acknowledges fruitful discussions with K. Ishida and M. Sato. H. Leng acknowledges the Chinese Scholarship Council for grant 201604910855.

\end{acknowledgements}

\bibliography{References_PdTe2}

\bibliographystyle{apsrev4-1}

\end{document}